\documentclass[prl,twocolumn,amsmath,amssymb]{revtex4-1} %

\usepackage{epsfig,amsmath}
\usepackage{subfigure}
\usepackage{graphicx}
\usepackage{dcolumn}
\usepackage{stmaryrd}
\usepackage{mathrsfs}
\usepackage{pifont}
\usepackage{amsthm}
\usepackage{amssymb}
\usepackage{bm}
\usepackage{latexsym}
\usepackage{hyperref}
\usepackage{color}

\newcommand{\la}{\langle}
\newcommand{\ra}{\rangle}
\newcommand{\lam}{\lambda}
\newcommand{\ti}{\tilde}
\newcommand{\ga}{\gamma}
\newcommand{\Ga}{\Gamma}

\newcommand{\da}{\dagger}
\newcommand{\De}{\Delta}

\newcommand{\om}{\omega}
\newcommand{\Om}{\Omega}

\newcommand{\pa}{\partial}

\def\pra#1{{ Phys.\ Rev. A\/} {\bf#1}}
\def\prb#1{{ Phys.\ Rev. B\/} {\bf#1}}

\def\prl#1{{ Phys.\ Rev.\ Lett.} {\bf#1}}

\def\pla#1{{ Phys.\ Lett. A\/} {\bf#1}}

\def\jcp#1{{ J.\ Chem. \ Phys} {\bf#1}}

\begin{document}

\title{Nonperturbative Leakage Elimination Operators and Control of a Three-Level System}

\author{Jun Jing$^{1,2}$, Lian-Ao Wu$^{2,3}$\footnote{Author to whom any correspondence should be addressed. Email address: lianao.wu@ehu.es}, Mark Byrd$^{4}$, J. Q. You$^{5}$, Ting Yu$^{6}$, Zhao-Ming Wang$^{7}$}

\affiliation{$^{1}$Institute of Atomic and Molecular Physics and Jilin Provincial Key Laboratory of Applied Atomic and Molecular Spectroscopy, Jilin University, Chuangchun 130012, Jilin, China \\ $^{2}$Department of Theoretical Physics and History of Science, The Basque Country University (EHU/UPV), PO Box 644, 48080 Bilbao, Spain \\ $^{3}$Ikerbasque, Basque Foundation for Science, 48011 Bilbao, Spain \\ $^{4}$Physics Department, Southern Illinois University, Carbondale, Illinois 62901-4401, USA \\ $^{5}$ Beijing Computational Science Research Center, Beijing 100084, China \\ $^{6}$ Center for Controlled Quantum Systems and Department of Physics and Engineering Physics, Stevens Institute of Technology, Hoboken, New Jersey 07030, USA, \\$^{7}$ Department of Physics, Ocean University of China, Qingdao 266100, China}

\date{\today}

\begin{abstract}
Dynamical decoupling operations have been shown to reduce errors in quantum information processing.  Leakage from an encoded subspace to the rest of the system space is a particularly serious problem for which leakage elimination operators (LEO) were introduced.  These are a particular type of decoupling which are designed to eliminate such leakage errors.  Here, we provide an analysis of non-ideal pulses, rather than the well-understood ideal pulses or bang-bang controls.  We show that under realistic conditions for experiments these controls will provide protection from errors. Furthermore, we find that the effect of LEOs depends exclusively on the integral of the pulse sequence in the time domain with proper ratio of pulse duration time and its period. When these two key parameters are chosen within certain bounds, leakage errors of the open system (exemplified by a three-level system for the nitrogen-vacancy centers under external magnetic field) would be dramatically decreased. The results are illustrated by the fidelity dynamics of LEO sequences, ranging from regular rectangular pulses, random pulses and even disordered (noisy) pulses.  
\end{abstract}


\maketitle

{\em Introduction.}---Leakage from a subspace encoding a qubit into the larger Hilbert space of a system's Hilbert space is particularly damaging since it removes any benefit the encoding may provide.  Leakage elimination operators (LEOs) were originally proposed to counteract the influence of leakage operators (denoted $L$) in a two-level system which encodes one logical qubit in a multi-level Hilbert space \cite{LEO,LEO2,Ari1,Ad1}. The leakage elimination was achieved by employing unbounded fast and strong ``bang-bang'' (BB) pulses \cite{BB} which apply to first order corrections of the evolution. In general, the total Hamiltonian for system and bath can be written as $H_{\rm SB}=H_P+H_Q+H_L$, where the operator of type $P$ represents the operations only acting within the qubit subspace, i.e., the subspace of interest. The operator of type $Q$ has no effect on the qubit subspace because it acts only within the remaining subspace of the whole Hilbert space perpendicular to the subspace of $P$, and $L$ represents the diffusion between the $P$- and $Q$-subspaces \cite{PQ}. The leakage error induced by the inevitable decoherence or diffusion can ruin the open system state by invalidating the encoding of qubits. If an operator $R_L$ satisifies $\{R_L, L\}=0$ and $[R_L, P]=[R_L, Q]=0$, then it follows that $R_L$ is a leakage elimination operator: $\lim_{m\rightarrow\infty}(e^{-i\frac{H_{\rm SB}t}{m}}R^\da_Le^{-i\frac{H_{\rm SB}t}{m}}R_L)^m=e^{-iH_Pt}e^{-iH_Qt}$. This holds to order of $t^2$ when $m=1$.

For this to be a good approximation, the pulse described by $R_L$ or $R^\da_L$ is so strong and fast that the system-bath Hamiltonian can be effectively turned off during these BB control seqences. (\cite{Lidar} and references therein.) This assumption is impractical or almost experimentally inaccessible for most existing setups. Another defect  is the choice of free evolution time $t$. For BB pulses, one should make $t\ll1/\om_c$, where $\om_c$ is the upper bound of the bath characteristic frequency for those modes coupling to the system. It is reasonable for an open system in its environment, where the characteristic frequency of the environment is much less than that of the system. When the frequencies of both system and environment (the latter can be inferred from the environmental memory time) are comparable to each other, it is difficult to satisfy these requirements.  

Therefore, a nonperturbative version of LEO theory is desirable for the coherence-protection/diffusion-suppression protocol for open quantum system.  This would enable the use of these sequences in wider domain of the system's characteristic parameters, which would apply for non-ideal pulses 
in the presence of a non-Markovian environment. A practical example is the effective three-level Hamiltonian for the spin of electronic ground state of a nitrogen vacancy (NV) center \cite{NV1} in a diamond crystal in the presence of an external magnetic field. An NV center has an $S=1$ state with zero-field splitting $D=2.88$ GHz between the $m_s=0$ and $m_s=\pm 1$ states. An external magnetic field along the crystalline axis of the diamond will lift the degeneracy of the $m_s=\pm 1$ states. The lowest two levels with $m=0$ and $m=-1$ have an energy gap $\om_{\rm NV}=D-g_e\mu_BB_z\approx(2.88-0.1B_z/mT)$ GHz and it can be used as the spin-based quantum memory unit \cite{NV2}. However, this neglects the influence of state $m_s=1$. The fluctuation of the external magnetic field would violate the far off-resonance condition between states $m_s=1$ and $m_s=-1$. Spin-echo sequences incorporated into the gate operation for removing the low-frequency noise during single-qubit evolution will affect the superposition state between $m_s=0$ and $-1$. Leakage problems can be serious especially in a small magnetic field.  

In this work, the theory of nonperturbative LEO is presented in the framework of non-Markovian quantum-state-diffusion (QSD) equation \cite{QSD}, by which an arbitrary sequence of LEO pulses as well as their fluctuations can be taken into account during the time evolution of the open system without any approximation. In particular, we focus on the leakage problem of three-level system that is universal in the quantum optics and quantum solid-state devices. Yet our protocol allows for a straightforward extension to systems with an arbitrary number of levels. Distinguished from other work targeting the optimized pulse sequence along the line of BB control \cite{UDD}, below we use the regular, random and noisy pulse sequences \cite{NPDD} to identify the key elements or parameters for attaining decoherence-suppression in addition to demonstrating the control dynamics under LEO.  This is done numerically, not perturbatively.  

{\em Construction of nonperturbative LEO.}---Rather than the additional BB pulse interrupting the free evolution under the total Hamiltonian $H_{\rm SB}$, the LEO here constitutes one part of system Hamiltonian in QSD equation. The total system space is separated into the subspace in which we are interested, the $P$-subspace, and the remaining $Q$-subspace.  The LEO acts as $I$ in $P$ and $-I$ in $Q$, i.e., $R_L={\rm diag}[c_1I, -c_2I]$, where the two identity operators have the same dimensions as $P$ and $Q$, respectively; and $c_{k}$'s ($k=1,2$) are c-numbers. Note that this is different from the ideal BB case presented in \cite{LEO2}.

Consider a general three-level atomic system \cite{threelevel}: $H_{\rm sys}=\sum_{j=1}^3\om_j|j\ra\la j|$. The Lindblad operators for the $V$-type and $\lambda$-type atoms are denoted by $L_V=\mu_1|3\ra\la1|+\mu_2|3\ra\la2|$ and $L_{\lam}=\nu_3|3\ra\la1|+\nu_2|2\ra\la1|$, respectively. Then by the LEO protocol, the nonperturbative operators can be written as
\begin{equation}
R_L^V=c(t)\left(\begin{array}{ccc} 1 & 0 & 0 \\  0 & 1 & 0 \\ 0 & 0 & 0 \end{array}\right), \quad R_L^\lam=c(t)\left(\begin{array}{ccc} 1 & 0 & 0 \\  0 & 0 & 0 \\ 0 & 0 & 0 \end{array}\right),
\end{equation}
where $c(t)$ is the implemented pulse sequence. By the QSD equation \cite{QSD}, the stochastic wave-function for the system including the LEO is governed by the following equation of motion (setting $\hbar=1$):
\begin{equation}\label{SS}
\pa_t\psi_t(z^*)=[-iH_{\rm sys}-iR_L^x+L_xz_t^*-L_x^\da\bar{O}_x(t)]\psi_t(z^*),
\end{equation}
where $x=V$ or $\lam$.

For a $V$-type three-level atom, $\bar{O}_V(t)=F_1(t)|3\ra\la1|+F_2(t)|3\ra\la2|$, where $F_k(t)\equiv\int_0^tds\alpha(t,s)f_k(t,s)$, $k=1,2$. $\alpha(t,s)$ is the environmental correlation function and $f_k(t,s)$ satisfies $f_k(t,t)=\mu_k$ and $\pa_tf_k(t,s)=i[\om_k-\om_3+c(t)]f_k+F_k(t)(\mu_1f_1+\mu_2f_2)$. While for the $\lam$-type system, $\bar{O}_\lam(t)=P_2(t)|2\ra\la1|+P_3(t)|3\ra\la1|$, where $P_k(t)\equiv\int_0^tds\alpha(t,s)p_k(t,s)$, $k=2,3$, and $p_k(t,s)$ satisfies $p_k(t,t)=\nu_k$ and $\pa_tp_k(t,s)=i[\om_1-\om_k+c(t)+\nu_2P_2+\nu_3P_3]p_k$. According to Eq.~(\ref{SS}), the ansatz $\bar{O}_x$, and the Novikov theorem, the exact master equation in the rotating frame with respect to $H_{\rm sys}+R_L^x$ is 
\begin{equation}\label{decoh}
\pa_t\rho_{\rm sys}=[L_x, \rho_{sys}\bar{O}_x^\da]+[\bar{O}_x\rho_{sys}, L_x^\da].
\end{equation}
The fidelity describing the survival probability of the initial state $\psi_0$ is defined by $\mathcal{F}(t)\equiv\sqrt{\la\psi_0|\rho_{\rm sys}|\psi_0\ra}=\sqrt{M[\la|\psi_0|\psi_t(z^*)\ra\la\psi_t(z^*)|\psi_0\ra]}$, where $M[\cdot]$ indicates an ensemble average.

\begin{figure}[htbp]
\centering
\subfigure{\label{diagram}
\includegraphics[width=2.8in]{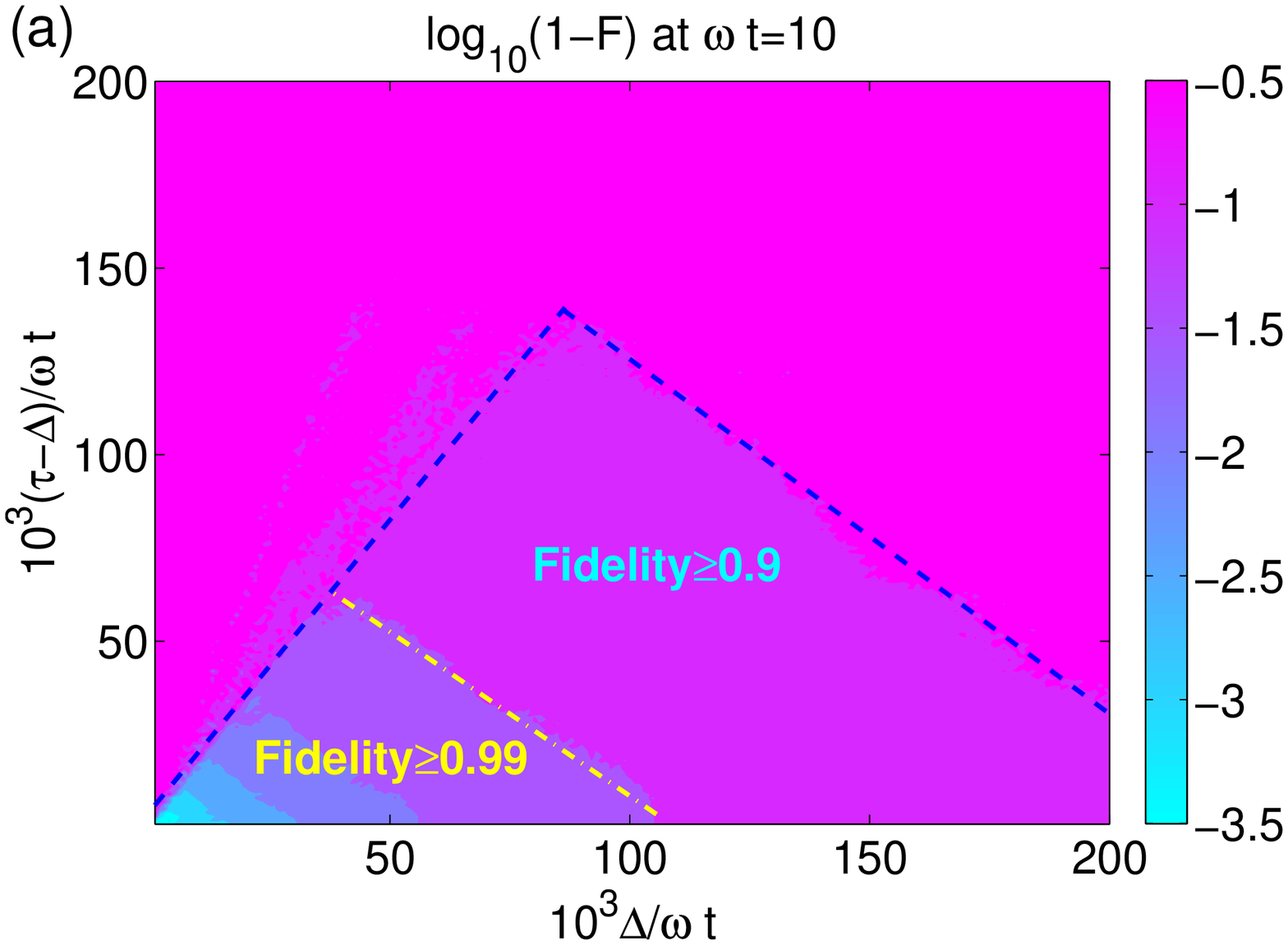}}
\subfigure{\label{Phi0}
\includegraphics[width=2.8in]{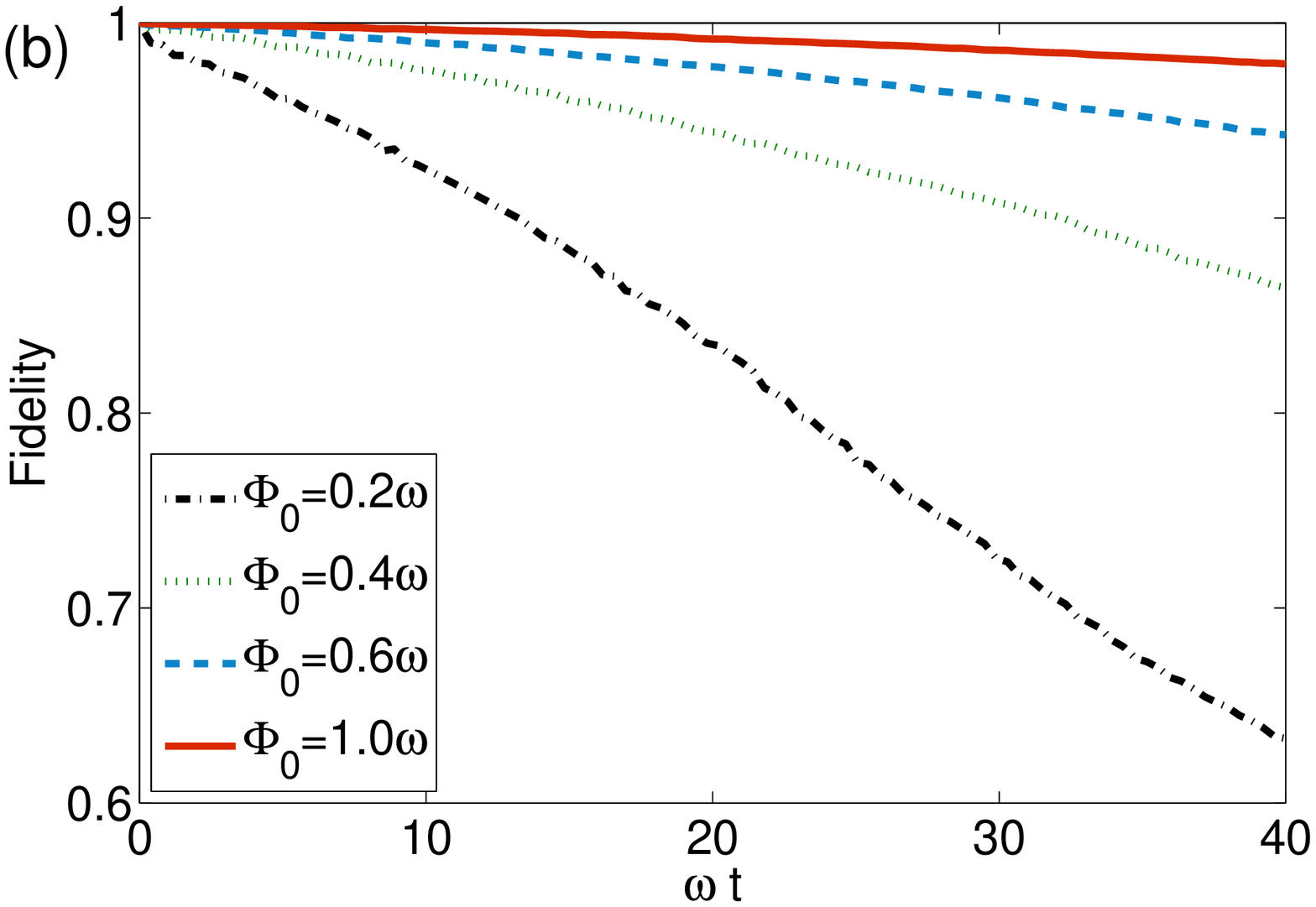}}
\caption{(Color online) (a) If one takes $|\psi_0\ra=|1\ra$ for the $\lam$-type system with $\om_1=\om/2$, $\om_2=\om_3=-\om/2$ and $|\psi_0\ra=1/\sqrt{2}(|1\ra+|2\ra)$ for the $V$-type atom with $\om_1=\om_2=\om/2$, $\om_3=-\om/2$, one can show that the expression for the fidelity is the same for both. This is the parameter diagram for fidelity at $\om t=10$ for these two typical states. The strength parameter is taken to be $\Phi_0=\om$. (b) The dynamics of a $V$-type three-level system under a regular LEO sequence with different strengths $\Phi_0$ for $|\psi_0\ra=1/\sqrt{2}(|1\ra+|2\ra)$. $\De/\tau=0.6$, $\tau=0.02\om t$. In both figures, the correlation function of the environment is $\alpha(t,s)=\Ga\ga/2e^{-\ga|t-s|-i\Om(t-s)}$, where the parameters are chosen to be $\ga=1$, $\Ga=\om$, and $\Om=0.5\om$.}
\end{figure}

{\em Result of nonperturbative LEO.}---A BB pulse is a limiting case, i.e., an approximation to a more practical rectangular pulse. The latter can be characterized by three parameters: the period $\tau$, the duration time $\De$, and the strength $\Phi_0$. In particular, $c(t)=\Phi_0/\De$ for $n\tau-\De\leq t\leq n\tau$, $n\geq1$ is an integer; otherwise, $c(t)=0$. In Fig. \ref{diagram}, we demonstrate a typical parameter diagram for the fidelity of a three-level atom coupled to a non-Markovian environment with an exponential decay correlation function, where $1/\ga$ characterizes the environmental memory time. Here the LEO control is chosen to be a rectangular pulse sequence and the y-axis denotes the dark time in one period of pulse. This diagram shows the region parameterized by $\De$ and $\tau$, where the fidelity can preserved at $0.9$, $0.99$, etc. It illustrates that the BB pulse merely occupies the lower left corner of the diagram and one can expect a tolerance from non-ideal pulses which achieve the same fidelity at any desired moment. It means one has a great deal of freedom to select an efficient combination of duration time and period. E.g., $\mathcal{F}\geq0.99$ can be obtained at $\om t=10$ as long as the ratio of dark time and duration time is not larger than about $3/2$ when $\De\leq0.04\om t$.

This result naturally raises a question: {\it what are the important parameters for attaining nearly the same control effect besides the parameters of environment?} Through numerical simulation over all of the parameters of LEO pulse, it turns out that within a fixed evolution time scale, the time integral over the pulse, i.e., the accumulation of the pulse strength in the control history, should be close to the ideal one. It is also clear that the pulse integral over the same time is linearly proportional to $\Phi_0$. In Fig. \ref{Phi0}, we compare the dynamics of a $V$-type three-level system under control with different $\Phi_0$. The calculations are performed with $\om_1-\om_3=\om$, $\om_2-\om_3=0.8\om$, $\mu_1=\om$, $\mu_2=0.5\om$. It is shown that the larger $\Phi_0$, the more robust the system is to the decoherence and leakage. E.g., at the fixed moment $\om t=40$, when $\Phi_0=0.4\om$, the fidelity decays to about $0.85$; while when $\Phi_0=\om$, the fidelity could be still above $0.98$.

\begin{figure}[htbp]
\centering
\subfigure{\label{Reg}
\includegraphics[width=2.8in]{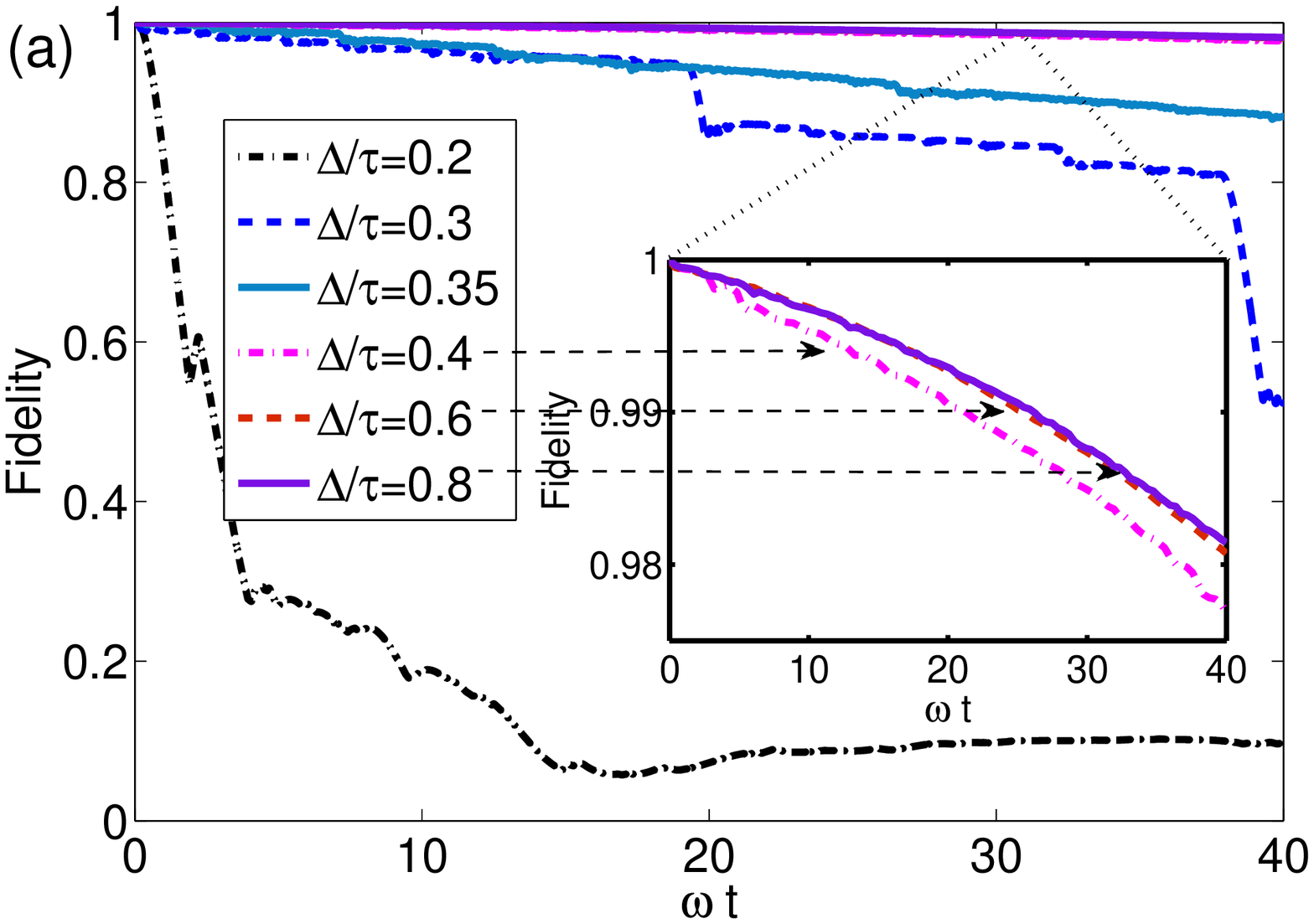}}
\subfigure{\label{ran}
\includegraphics[width=2.8in]{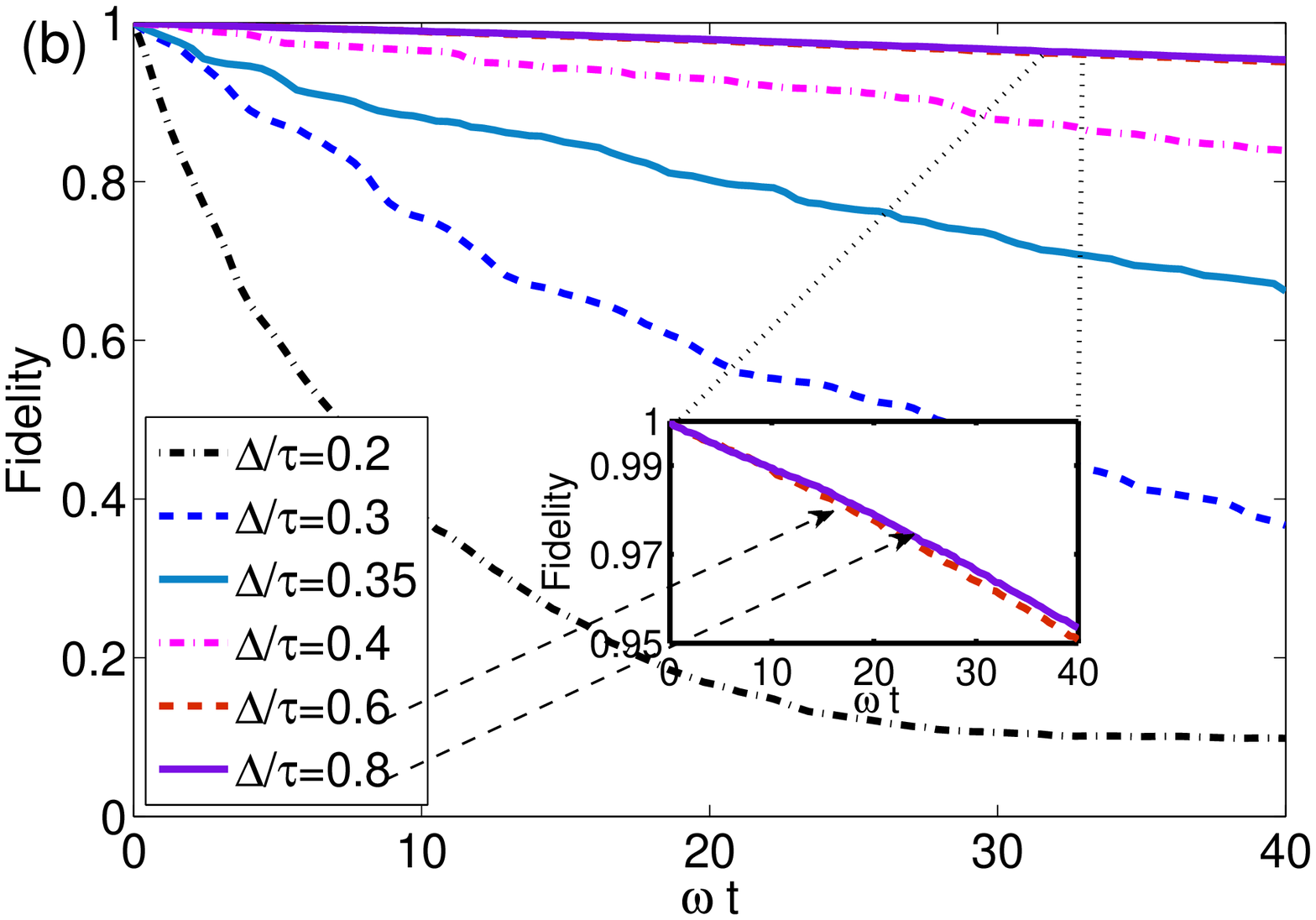}}
\caption{Dynamics of a $V$-type three-level system under (a) regular LEO and with different ratio $\De/\tau$ (b) random LEO with different ``average ratio'' $\De/\tau$. $|\psi_0\ra=1/\sqrt{2}(|1\ra+|2\ra)$. The environmental conditions are chosen to be the same as in Figs. \ref{diagram} or \ref{Phi0}.}
\end{figure}

However, this does not completely determine the key elements for the most effective LEO control. Fig. \ref{Reg} shows a numerical simulation that is performed with an unchanged pulse strength $\Phi_0$ and various ratios of duration time and period. It exhibits a phenomenon somewhat analogous to a phase transition, where $r\equiv\De/\tau=0.35$ seems to be the value of a threshold $r_c$. If $\De/\tau<r_c=0.35$, then accelerated decoherence process occurs in the dynamics of fidelity. The line of $\De/\tau=0.2$ shows a much more rapid decay in a very short time. Although when $\om t<19$, the control effect of $\De/\tau=0.3$ is almost the same as that of $\De/\tau=0.35$, it deviates from the asymptotic curve by several sudden jumps. If $\De/\tau=0.35$, then the fidelity decays to sightly less than $0.9$ at $\om t=40$, and shows a significant gap between this and results with even larger $\De/\tau$. While if $\De/\tau>r_c$, almost the same degrees of decoherence-suppression are achieved. See the insert in Fig. \ref{Reg} which shows the curves for $\De/\tau=0.4, 0.6, 0.8$ have a fidelity approximately $0.98$ at $\om t=40$, and the maximal relative error is less than $0.5\%$. So here the fidelity will be saturated in the regime of $0.4<\De/\tau<1$, where the control effect of the LEO is almost completely determined by the pulse integral over time rather its configuration.
s

The duration time and period to produce a regular rectangular pulse might not be obtainable through the  available controls in practice. One can never completely eliminate stochastic quantum fluctuations and environmental noise which inevitably yield random rather than regular pulses. Such a random sequence can be ``simulated'' in the following way: based on a regular sequence with fixed $\tau$, $\De$ and $\Phi_0$, the time-dependent quasi-period and pulse strength are determined by $X'=X[1+A_XR_X(t)]$, where $X=\tau$ or $\Phi_0$, $R_X$ can be uniformly distributed between $-1$ and $1$, and $R_\tau$ and $R_{\Phi_0}$ are uncorrelated. After a sufficiently long evolution time and ensemble average, $M[X']=X$. Therefore the integral of the pulse strength over time should be the same as that of the original regular sequence. Figure \ref{ran} shows the result after ensemble averaging, where random amplitudes are chosen to be between $A_\tau=40\%$ and $A_{\Phi_0}=90\%$. Compared to Fig.~\ref{Reg}, accelerated-decoherence phenomenon has been reduced. However, under such a large fluctuation in the parameters, the effect of the controls with the same $\De/\tau\leq0.4$ seems to be only a little less effective than that of the regular LEO pulse. One can still find the asymptotic  fidelity saturation when $\De/\tau\geq0.6$ [See the insert of Fig.~\ref{ran}.], by which the fidelity decays to around $0.96$ at $\om t=40$.

\begin{figure}[htbp]
\centering
\includegraphics[width=2.8in]{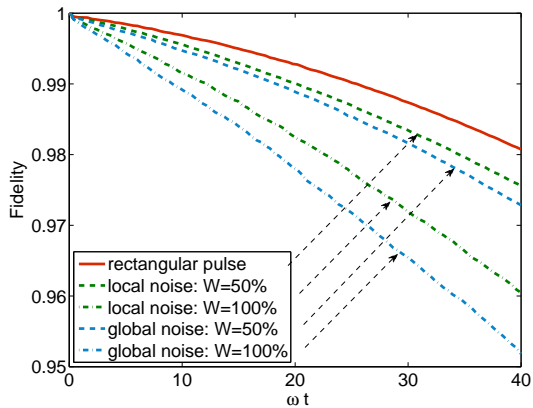}
\caption{Dynamics of a $V$-type three-level system under global (blue lines) and local (green lines) noisy LEOs with different amplitudes of Gaussian noise. $|\psi_0\ra=1/\sqrt{2}(|1\ra+|2\ra)$, $\De/\tau=0.6$, $\tau=0.02\om t$. The environmental conditions are chosen to be the same as that in Fig. \ref{diagram} or \ref{Phi0}.} \label{fluc}
\end{figure}


Under certain circumstances, a real nonperturbative LEO control can be a control of noisy pulse influenced by some out-of-control environmental factor. If the integral over pulse strength during the same period of time is to be unchanged, one can expect two types of noises. (a) Global noise: $c(t)\rightarrow c(t)+\Phi_0/\tau Wn(t)$, where $W$ is a percentage measuring the dimensionless noise strength and $n(t)$ is a white noise with uniform or Gaussian distribution; (b) Local noise: only in each duration time $\De$  the strength of pulse becomes the fixed value plus the noise $\Phi_0/\tau Wn(t)$ while it remains dark during the dark intervals. It turns out that with the global uniformly distributed white noise, the fidelity dynamics of the system is nearly the same as that under the regular pulse sequence. The results of Gaussian noise are presented in Fig. \ref{fluc}. In order to exhibit the effect of the pulse integral, we use a greatly exaggerated amplitude of the noise. It is found that its deviation from the regular pulse (see the values at the last time point in Fig. \ref{fluc}) is slightly increased by $W$. Yet even when $W$ is as large as $100\%$, for local noise, the deviation is still less than $0.02$ ($\approx0.98-0.96$); for global noise, it is less than $0.03$. Therefore it is reasonable to suppose that the accumulation of pulse strength remains the most important element even for a noisy LEO pulse sequence.

{\em Discussion}---The idea of nonperturbative LEO is to rapidly rotate the quantum system by means of control fields to average the system-environment coupling to zero. More particularly, under an LEO, the modulus of coefficient functions $F_k$'s, $k=1,2$, of O-operator in the master equation (\ref{decoh}) that determines the rate of decoherence can be preserved as close as possible to their boundary value $0$. Using the correlation function indicated in Fig. \ref{diagram}, one can find their equations of motion $\dot{F}_k=ic(t)F_k+G(t)$, where $G(t)=\frac{\Ga\ga\mu_k}{2}+[-
\ga+i(\om_k-\om_3-\Om)]F_k+(\mu_1F_1+\mu_2F_2)F_k$. Set $\ti{F}_k=e^{-i\int_0^tdsc(s)}F_k\equiv e^{-iC(t)}F_k$.  Then $\dot{\ti{F}}_k(t)=e^{-iC(t)}G(t)$, and their formal solutions are
\begin{equation}\label{Ct}
\ti{F}_k(t)=\int_0^tdse^{-iC(s)}G(s).
\end{equation}
When $C(t)$ (the integral over pulse) is sufficiently large, the kernel in the integral of Eq.~(\ref{Ct}) consists of a fast-oscillation function $e^{-iC(t)}$ and a slowly variation function $G(t)$, so that $\ti{F}_k$ vanishes \cite{NPDD}. Obviously, the first term in $G(t)$ is a constant; the second one is the linear term of $F_k$, the modulus of the  coefficient is $\sqrt{\ga^2+(\om_k-\om_3-\Om)^2}$; and the last term is proportional to the square of $F_k$, which can be ignored if the variation of $F_k$ is small.  [I.e. it is consistent with a vanishing $F_k$ under the condition of large $C(t)$]. Thus one can also understand the reason that nonperturbative dynamical decoupling works well with small $\ga$ corresponding to a strong non-Markovian regime for environment. Although, the exponential correlation function we used here is not general, some other correlation functions, e.g., the $1/f$ noise, could be decomposed into a finite summation of this form with different $\ga$'s and $\Om$'s \cite{Zhou}. So the corresponding analysis would be quite similar to this model and our results could be readily extended to many other types of noise.  

Here our LEO can be used to protect the subspace of the NV spin against fluctuations of the magnetic field in presence of low field. Comparing to the previous dynamical decoupling prescription \cite{UDD}, the LEO presented here can not only obtain a larger coherence time, but also can reduce leakage of the effective qubit space, i.e. it can suppress the diffusion of the $m_s=\pm 1$ states as long as one performs a nonperturbative control field with compatible frequency on the order of $\om_{\rm NV}$. The results shown in Fig. \ref{Phi0} appy even in the presence of a much larger  dissipation coupling than is often met in practice.  (This is indicated by $\Ga\sim\om_{\rm NV}$, on the order of GHz, while in practical situations, it is usually less than $50$ MHz.)  In these cases, the fidelity of the system is still maintained as large as $0.98$ after $40$ ns, which is already much larger than a typical quantum-storage-operation time $\sim10$ ns.

{\em Conclusion}---In this letter, we presented a nonperturbative LEO approach to dynamical decoupling of an arbitrary multi-level system under the influence of a non-Markovian environment.  Based on the QSD equation, as well as the exact master equation, we found that the integral over the pulse sequence is the most important quantity in determining the effect of decoherence suppression under the condition of proper average ratio of pseudo-duration time to pseudo-period.  This result is not sensitive to fluctuations of the pulse strength and period and robustly removes disturbances of the system due to environmental noise under imperfect and noisy controls.

{\em Acknowledgments.}---We acknowledge grant support from the Basque Government (grant IT472-10), the Spanish MICINN (No. FIS2012-36673-C03-03), the NBRPC No.~2014CB921401, the NSAF No. U1330201, Science and Technology Development Program of Jilin Province of China(20150519021JH), and partially from the NSFC No. 11475160, the NSF No. PHY-0925174 and the AFOSR No. FA9550-12-1-0001.

\end{document}